\documentstyle[12pt,html,epsf,psfig]{article}

\begin{document}
\title{MATTERS OF GRAVITY, The newsletter of the APS Topical Group on 
Gravitation}
\begin{center}
{ \Large {\bf MATTERS OF GRAVITY}}\\ 
\bigskip
\hrule
\medskip
{The newsletter of the Topical Group on Gravitation of the American Physical 
Society}\\
\medskip
{\bf Number 26 \hfill Fall 2005}
\end{center}
\begin{flushleft}

\tableofcontents
\vfill
\section*{\noindent  Editor\hfill}

Jorge Pullin\\
\smallskip
Department of Physics and Astronomy\\
Louisiana State University\\
Baton Rouge, LA 70803-4001\\
Phone/Fax: (225)578-0464\\
Internet: 
\htmladdnormallink{\protect {\tt{pullin-at-lsu.edu}}}
{mailto:pullin@lsu.edu}\\
WWW: \htmladdnormallink{\protect {\tt{http://www.phys.lsu.edu/faculty/pullin}}}
{http://www.phys.lsu.edu/faculty/pullin}\\
\hfill ISSN: 1527-3431
\begin{rawhtml}
<P>
<BR><HR><P>
\end{rawhtml}
\end{flushleft}
\pagebreak
\section*{Editorial}

With this newsletter I would like to announce my retirement as editor.
I will edit one more newsletter (the February one) but from fall 2006
editorship will be taken over by David Garfinkle, as per decision of
the GGR executive committee.  It has been 11 years since I started
editing the newsletter and it was time for a change. We all wish David
great success with the editorship.

Some news from the executive committee: we will put on the ballot a
modification of the GGR bylaws to create the post of membership
coordinator, as our membership continues to grow (we are now at 782
members!). We will also propose an amendment of a sloppy wording about
the constitution of the nominating committee. The current bylaws state
it will have three members and then go on to describe how to choose
four (!)  members. This year the committee is chaired by Bei-Lok Hu,
helped by Joe Giaime (APS representative), Don Marolf and Nergis
Mavalvala.

The next newsletter is due February 1st. All issues are available in
the WWW:\\\htmladdnormallink{\protect
  {\tt{http://www.phys.lsu.edu/mog}}} {http://www.phys.lsu.edu/mog}\\
The newsletter is available for Palm Pilots, Palm PC's and web-enabled
cell phones as an Avantgo channel. Check out
\htmladdnormallink{\protect {\tt{http://www.avantgo.com}}}
{http://www.avantgo.com} under technology$\rightarrow$science.  A
hardcopy of the newsletter is distributed free of charge to the
members of the APS Topical Group on Gravitation upon request (the
default distribution form is via the web) to the secretary of the
Topical Group.  It is considered a lack of etiquette to ask me to mail
you hard copies of the newsletter unless you have exhausted all your
resources to get your copy otherwise.

If you think a topic should be covered by the newsletter you are
strongly encouraged to contact the relevant correspondent.  If you
have comments/questions/complaints about the newsletter email me. Have
fun.

\hfill Jorge Pullin
\vfill\eject

\bigbreak

\vspace{-0.8cm}
\parskip=0pt
\section*{Correspondents of Matters of Gravity}
\begin{itemize}
\setlength{\itemsep}{-5pt}
\setlength{\parsep}{0pt}
\item John Friedman and Kip Thorne: Relativistic Astrophysics,
\item Bei-Lok Hu: Quantum Cosmology and Related Topics
\item Gary Horowitz: Interface with Mathematical High Energy Physics and
String Theory
\item Beverly Berger: News from NSF
\item Richard Matzner: Numerical Relativity
\item Abhay Ashtekar and Ted Newman: Mathematical Relativity
\item Bernie Schutz: News From Europe
\item Lee Smolin: Quantum Gravity
\item Cliff Will: Confrontation of Theory with Experiment
\item Peter Bender: Space Experiments
\item Jens Gundlach: Laboratory Experiments
\item Warren Johnson: Resonant Mass Gravitational Wave Detectors
\item David Shoemaker: LIGO Project
\item Peter Saulson: former editor, correspondent at large.
\end{itemize}
\section*{Topical Group in Gravitation (GGR) Authorities}
Chair: Jorge Pullin; Chair-Elect: \'{E}anna Flanagan; Vice-Chair: 
Dieter Brill. 
Secretary-Treasurer: Vern Sandberg; Past Chair: Jim Isenberg;
Delegates:
Bei-Lok Hu, Sean Carroll,
Bernd Bruegmann, Don Marolf, 
Vicky Kalogera, Steve Penn.
\parskip=10pt

\vfill
\pagebreak

\section*{\centerline {
The WYP speakers program}}
\addtocontents{toc}{\protect\medskip}
\addtocontents{toc}{\bf GGR News:}
\addcontentsline{toc}{subsubsection}{\it
The WYP speakers program, by Richard Price}
\begin{center}
Richard Price, University of Texas at Brownsville
\htmladdnormallink{rprice-at-phys.utb.edu}
{mailto:rprice@phys.utb.edu}
\end{center}

In the spring of 2002 the Topical Group, and Jim Isenberg in
particular, calculated that 2005 would be coming, and with it the
World Year of Physics. To celebrate and exploit WYP, the Executive
Committee approved a plan to sponsor a WYP Speakers Program. A task
force (Jim Isenberg, Jennie Traschen, along with me as chair) was
appointed, and a clear structure and philosophy were identified. The
TGG role would be to assemble a list of speakers who were known to be
good at outreach, and would try to find a speaker from that list to
meet requests for WYP speakers.  From the very outset, the
prioritization of filling requests was meant to be based on TGG
self-interest; highest priority was to be given to requests from
colleges and universities that might supply graduate students to
gravity research groups.

The program has been housed at my new academic institution, the Center
of Gravitational Wave Astronomy (CGWA) at UT-Brownsville, with the
heavy housework done by staff member Danka Mogilska. When a request is
received Danka tries to locate a speaker appropriate for the
request. What is most appropriate is a speaker living a short driving
distance from the site of the request, since the program has had no
travel funds. Requests have come almost exclusively through a website

\htmladdnormallink{http://www.phys.utb.edu/WYPspeakers/REQUESTS/howto.html}
{http://www.phys.utb.edu/WYPspeakers/REQUESTS/howto.html}

The APS central organization would not offer any direct assistance.
Then and now, they did not consider the Speakers Program to be ``APS
wide." The APS and the AAPT did help by linking the Speakers Program
to their websites. The funding (support for Danka Mogilska) for the
program to this point has come from the CGWA.

The program seems to have been a success. We have met about half of
about 200 requests for the first half of 2005.  A second part of the
program was not a success. There was to have been a speakers database,
a compilation of talks, images, links, etc. used by speakers, that
would ease the preparation of a new presentation. Despite some
prodding of speakers there have been almost no contributions to this
database, and it will be abandoned.

Several months into the program we were joined by the Forum on the
History of Physics (FHP), represented by Virginia Trimble. Virginia
also served as a link to the Division of Astrophysics. Speakers from
the Forum were particularly useful when groups specifically requested
talks about historical and philosophical aspects of Einstein's
science. Furthermore, FHP/DAP had travel money for especially
interesting outreach opportunities.  For the past several months the
program has run as a loose collaboration of our original program and
Virginia's efforts.

The response to the program does indicate that there is a market for
the original program. Our 200 requests for the first half of 2005 may
not be a good measure of what the program is going to be. A message
from the APS recently went out to physics department chairs and we
have been swamped as a result: 50 requests in the last week.  The
requests also show that there is a market out there for talks to high
schools, nonacademic organizations, etc., a market we did not focus
on. So the 2005 program has taught us some us some interesting lessons,
but 2005 is coming to and end.

At their most recent meeting, the TGG Executive Committee approved, in
principle, the continuation of the program. The CGWA can no longer
support the program, so a source of funding is needed.  For the last
few months Virginia Trimble and I have been looking into long term
possibilities for the program, into short term holding patterns, and
have been looking for funding.

The good news is that a funds have been found, at least for the short
term, and the program will continue to be housed at the CGWA while
longer-term possibilities are explored.

Thanks to Virginia Trimble's efforts the major source of the funding
will be a private donor: Wayne Rosing.  Rosing, who has been described
in the media as "...  a legendary figure in the computer industry and
a keen astronomer...," is the first senior fellow in mathematical and
physical sciences at the University of California, Davis. Following
Rosing's request, the program will be called the Las Cumbres Speakers
Program, and will include some support for speaker travel.  Both TGG
and FHP will also be contributing some funds to the program to help
support travel. The nature of the program can be expected to shift
towards serving a broader set of requests, and the mode of operation
will shift away from reliance on a list, and will include more ad hoc
improvisational searching for the right speaker to meet a request.

The long term future of the speakers program is still under
discussion, a discussion that will greatly benefit from what will be a
very interesting year for the program.

\section*{\centerline {
We hear that...}}
\addcontentsline{toc}{subsubsection}{\it
We hear that..., by Jorge Pullin}
\begin{center}
Jorge Pullin, LSU
\htmladdnormallink{pullin-at-lsu.edu}
{mailto:pullin@lsu.edu}
\end{center}

Hanno Sahlmann has won the 
2005 ``Akademiepreis f\"ur Physik" of the Gottingen Academy of 
Sciences. The citation says that the prize is being awarded 
for "important contributions to Loop Quantum Gravity, a promising 
framework for the quantization of Einstein's General Relativity." 

Abhay Ashtekar has been named  the Sir C.V. 
Raman Visiting Chair of the Indian Academy of Science and was awarded a 
senior Forschungspreis by the Humboldt Foundation

Hearty Congratulations!

\section*{\centerline {
100 Years ago}}
\addcontentsline{toc}{subsubsection}{\it  
100 Years ago, by Jorge Pullin}
\parskip=3pt
\begin{center}
Jorge Pullin
\htmladdnormallink{pullin-at-lsu.edu}
{mailto:pullin@lsu.edu}
\end{center}
An English version of 
``Does the inertia of a body depend on its energy content'' (aka ``The
$E=mc^2$ paper) is available 
at 
\htmladdnormallink{http://www.phys.lsu.edu/mog/100}
{http://www.phys.lsu.edu/mog/100}
\vfill
\eject

\section*{\centerline {
What's new in LIGO}}
\addtocontents{toc}{\protect\medskip}
\addtocontents{toc}{\bf Research Briefs:}
\addcontentsline{toc}{subsubsection}{\it  
What's new in LIGO, by David Shoemaker}
\begin{center}
David Shoemaker, LIGO-MIT
\htmladdnormallink{dhs-at-ligo.mit.edu}
{mailto:dhs@ligo.mit.edu}
\end{center}

There has been broad progress in LIGO over the last six months. One
first note on the Livingston Observatory, which was in the path of
Hurricane Katrina. The observatory staff are all safe. The site is
sufficiently inland that the worst of the wind and water damage was
avoided, although a building panel was ripped off, debris thrown
around, and the power lost for a period of time. At the time of
writing, it is being brought back to operational condition, and no
obvious significant damage was done.

At the time of the last MOG, we were anticipating the fourth science
run (S4). It took place as planned, and the bottom line is that the
detectors worked very well. They worked with high sensitivity and
duty cycle (thanks to the successful implementation of the hydraulic
`HEPI' pre-isolation system at Livingston). Diagnostic tools,
looking at the state of the instruments and the quality of the data,
were running in full force, giving feedback to the operators and
science monitors to watch for problems and allow tuning for the
highest quality of data.

A handful of papers from earlier science runs have now been
submitted. Since the last MOG, such bodice-rippers as ``Search for
gravitational waves from galactic and extra–galactic binary neutron
stars'', ``Search for Gravitational Waves from Primordial Black Hole
Binary Coalescences in the Galactic Halo'', ``Upper limits from the
LIGO and TAMA detectors on the rate of gravitational-wave bursts'',
``Upper limits on gravitational wave bursts in LIGO's second science
run'', and ``Upper Limits on a Stochastic Background of
Gravitational Waves'' have hit the gr-qc newstand. The last paper
puts the exciting upper limit of $\Omega_{gw}(f)<8.4\times 10^{-4}$
in the band of frequencies from 69-156 Hz, an improvement of some
$105$ in the limit at those frequencies. We are still catching up
on the data collected, and are working hard on the very nice recent
S4 data. Einstein@home, described in the last MOG, has been a
resounding success. It is providing roughly 20 Teraflops of
computing power and is working away at searches for periodic
sources, like pulsars, over broad sections of the sky. Please
consider donating your spare cycles to
http://einstein.phys.uwm.edu/.

Detector commissioning has continued with great success. On all
instruments, the core of the progress has been in increasing the
laser power. This pushes down the noise in the shot-noise limited
region, but equally importantly gives better signal-to-noise in many
auxiliary channels. In the case of the Livingston instrument, better
alignment control was needed, and achieved, to handle this power.
The 4km instrument at Hanford had shown excessive absorption in
several optics, and careful sleuthing showed that one of the input
optics of a 4km Fabry-Perot cavity was the prime suspect. A foray
into the vacuum system was made to switch out that optic, and to
clean another. This gave a factor of (at least) ten reduction in
absorption, paving the way for useful increases in power for this
instrument. The 2km instrument at Hanford also was tuned up, and
variety of other optical, mechanical, and controls aspects of the
interferometers have been addressed in parallel. The instruments
have now all performed effectively at their design sensitivity, and
the Collaboration has decided to proceed with the main initial LIGO
S5 Science Run, designed to collect a year of integrated data,
planned to start late this year.

Advanced LIGO design and prototyping continues apace. A full-size
prototype, from Caltech and UK design teams, of the quadruple test
mass suspensions has just been assembled. The seismic isolation
prototype at Stanford's test facility has demonstrated design
performance in many aspects, and we are now moving forward on the
prototype for the test mass chambers. Work has been done at MIT on
characterizing the seismic interface and at Caltech on alternative
designs for the auxiliary optics `HAM' chambers. Along with
technical progress on the laser at the Max-Planck Institute in
Hannover, funding for supplying the pre-stabilized lasers for
Advanced LIGO has now been assured. The good technical progress
allows us to be ready for and optimistic about a 2008 start of
funding for Advanced LIGO from the National Science Foundation.

One important evolutionary change is underway: The reorganization of
the LIGO Laboratory and the LIGO Scientific Collaboration (LSC). The
two are becoming one organization -- referred to simply as 'LIGO' --
as a recognition of the crucial role that the LSC plays in the
development and exploitation of the Observatories. Individuals from
the LSC are taking on central organizational roles, exemplified by
the addition of the LSC spokesperson to the leadership group of the
LIGO Director and Deputy Director. The LSC is also participating
fully in the search for a new Director, as Barry Barish has been
asked to take on the leadership of the global design effort for the
international linear collider. We will miss his leadership;
fortunately he will remain as an active member of the LSC. We are
also moving to bring our close international partners into the
organization; re-writing the LSC Charter and Bylaws; and generally
adjusting to the reality that we have a set of running detectors,
with a mature data analysis process, and an Advanced LIGO upgrade
which is starting to take on a life of its own.

\vfill
\eject
\section*{\centerline {Recent developments in the information loss paradox}}
\addtocontents{toc}{\protect\smallskip}
\addcontentsline{toc}{subsubsection}{\it
Recent developments in the information loss paradox, by \'Eanna Flanagan}
\begin{center}
    \'Eanna Flanagan, Cornell \\
\htmladdnormallink{eef3-at-cornell.edu}
{mailto:eef3@cornell.edu}
\end{center}
\parindent=0pt
\parskip=5pt

\def\alt{
\mathrel{\raise.3ex\hbox{$<$}\mkern-14mu\lower0.6ex\hbox{$\sim$}}
}

In 1976 Stephen Hawking noted that the formation and evaporation of
black holes as described in the semiclassical approximation appear to
transform pure states into mixed states, in conflict with unitarity
(Hawking 1976).   This apparent paradox engendered a lively debate
during the subsequent decades, with no generally accepted resolution.
A poll taking during a 1993 conference showed 32\% of the
participants for information loss, with the remainder in favor of
unitarity or other resolutions of the paradox (Page 1994).
If such a poll were repeated today, however, the result would probably
be markedly different --- evidence in favor of the preservation of
information has emerged from string theory, and more recently from
Euclidean quantum gravity (Hawking 2004,2005) and loop quantum gravity
(Ashtekar \& Bojowald 2005a,b).
There is thus an emerging consensus that black hole evaporation is
unitary.  However, the flaws in the original semiclassical arguments
have not yet been clearly resolved; this aspect of the subject remains
a puzzle.

The standard semiclassical argument for information loss is as follows
(see, eg, Preskill 1992, Strominger 1994).  Consider a solution of the
semiclassical Einstein equations that describes the formation and
evaporation of a black hole with initial mass $M \gg 1$ in Planck units.
It is argued that the domain
of validity of such a solution consists of all points in whose causal
pasts the local curvature is everywhere sub-Planckian; this belief is
supported by detailed calculations in two-dimensional models (Bose,
Parker \& Peleg 1996, Strominger 1994).  One can in principle excise
from the spacetime all points outside this domain of validity.
The resulting spacetime is generally believed to have the following
features (based again in part on explicit calculations in
two-dimensional models; see eg.\ Strominger 1994): (i) It contains an
apparent horizon whose area shrinks down to the Planck
scale due to Hawking emission. (ii) It can be foliated by spacelike slices
whose extrinsic curvatures are everywhere sub-Planckian such
that one of the slices, $\Sigma$, contains most of the outgoing
Hawking radiation (mass $\sim M$), yet
intersects the infalling matter inside the black hole
in a region of low curvature (Lowe et.\ al.\ 1995).

Next, one assumes that the black hole evaporates completely, and that
after the evaporation spacetime can again be described using the
semiclassical approximation.
The initial conditions for this final semiclassical spacetime region
are determined in part by quantum gravity.  One assumes that this region
can be foliated by spacelike slices, the first of which,
$\Sigma^\prime$, can be chosen to coincide with $\Sigma$ outside of
some closed two-surface $S$.  Thus, $\Sigma = \Sigma_{\rm in} \cup
\Sigma_{\rm out}$ and $\Sigma^\prime = \Sigma_{\rm in}^\prime \cup
\Sigma_{\rm out}$, where $\Sigma_{\rm out}$ is the region outside $S$,
$\Sigma_{\rm in}$ is the portion of $\Sigma$ inside $S$, which
intersects the infalling matter inside the black hole, and
$\Sigma_{\rm in}^\prime$ is the portion of $\Sigma^\prime$ inside $S$,
which contains the end-products of the last stages of evaporation.
The construction can be specialized so that the mass contained within
$S$ is of order the Planck mass (since the semiclassical description
of the evaporation is accurate down to the Planck scale), and so that
the radius $R$ of $S$ is of order the time taken for the black hole to
decay completely after it reaches the Planck scale (the ``remnant
lifetime'').

If the quantum state on $\Sigma^\prime$ were pure, then the
entanglement entropies $S_{\rm out} =
- {\rm tr}( {\hat \rho}_{\rm out} \ln {\hat \rho}_{\rm out})$
and $S_{\rm in}^\prime =- {\rm tr}( {\hat \rho}_{\rm in}^\prime \ln
{\hat \rho}_{\rm in}^\prime)$ of the density matrices on $\Sigma_{\rm
  in}^\prime$ and $\Sigma_{\rm out}$ would agree.  If the
Hawking radiation is approximated to be in a thermal state, its entropy is of
order $S_{\rm out} \sim M2$ (Page 1976).  One might
worry that this approximation is not reliable and that
subtle correlations between Hawking quanta could make the state on
$\Sigma_{\rm out}$ very nearly pure, giving $S_{\rm out} \ll M2$.
However since the state on $\Sigma$ is pure, $S_{\rm out} = S_{\rm
  in}$, so a nearly pure state on $\Sigma_{\rm out}$ would require a
nearly pure state on $\Sigma_{\rm in}$.  This would contradict the
linearity of quantum mechanical evolution from initial states of the
collapsing matter to states on $\Sigma$ (Preskill 1992).  Therefore
the estimate $S_{\rm out} \sim M2$ seems unavoidable.
Finally, if the lifetime $R$ is assumed to be of order the Planck
time, the region $\Sigma_{\rm in}^\prime$ is too small and contains
too little energy to allow significant correlations, $S_{\rm
  in}^\prime \alt 1$.  [Preskill (1992) shows that $S_{\rm in}^\prime
\sim M2$ would require a lifetime of order $R \sim M4$.]  Therefore
$S_{\rm in}^\prime \ll S_{\rm out}$ and the state on $\Sigma^\prime$
cannot be pure.

This semiclassical argument indicates that the information defined by $I = \ln {\rm dim}
{\cal H} + {\rm tr} ( {\hat \rho} \ln {\hat \rho})$ decreases, where
${\cal H}$ is the Hilbert space.  However, as pointed out by Page
(1994), it does not necessarily imply that information is lost in the
colloquial sense that knowledge of the final (mixed) state on $\Sigma^\prime$
is insufficient to compute the initial (pure) state: pure-to-mixed
evolutions can be described by
invertible superscattering matrices.  Nevertheless, if one accepts
that $I$ decreases, then information loss in the stronger, colloquial
sense seems inevitable, since the state on $\Sigma_{\rm out}$ as
computed in the semiclassical approximation depends only very weakly
on the initial quantum state of the infalling matter.

Many different resolutions to the information loss paradox have been
proposed; see eg.\ Thorlacius (2004) or Page (1994) for detailed
reviews.  Some novel recent suggestions are those of Horowitz \& Maldacena
(2004) and Gambini et.\ al.\ (2004).  In the remainder of this article I
will focus on evidence from various approaches to quantum gravity that
black hole evaporation is unitary.

The evidence from string theory is well-known; see the articles by
Gary Horowitz in issues 12 and 18 of Matters of Gravity.
Briefly, the AdS/CFT duality of string theory implies that the formation and
evaporation of small black holes in anti-deSitter space can described
completely (albeit indirectly) by a manifestly unitary conformal field theory.  String
theory also provides an explanation for the failure of the
semiclassical argument summarized above.  Namely, the low energy
effective theory describing evolution in the foliation used in the
semiclassical argument is {\it non-local} even in low-curvature
regimes: it contains states of strings stretched over macroscopic
distances (Lowe et.\ al.\ 1995, Lowe \& Thorlacius 1999).
This non-locality allows external observers to measure the information
coming out in the Hawking radiation, while infalling observers see
nothing unusual happening while the infalling matter crosses the horizon.
The idea of such low-energy non-locality is anathema to many relativists
(see, eg.\ Jacobson et.\ al.\ 2005), but is
supported by computations of commutators of spacelike separated
operators in string theory (Lowe et.\ al.\ 1995).  Thus, the key
assumption that semiclassical general relativity is a good
approximation at sub-Planckian curvatures is invalid in string theory.

Turn now to the Euclidean path integral approach to quantum gravity.
For many years, Stephen Hawking has been the most prominent proponent
of the view that black holes destroy information.  His dramatic and
much-publicized reversal on this issue last year at the GR17
conference (Hawking 2004) was described in Brien Nolan's article in
issue 24 of Matters of Gravity.  Hawking considers scattering
processes in anti-deSitter space that classically would be expected to
produce a black hole.  He argues in this context that
a path integral computation of correlation
functions of operators at infinity will be dominated by a sum over two
topological sectors, one corresponding to no black hole being present
and one corresponding to a black hole being present.
(Since observers making measurements at infinity cannot
be sure if a black hole formed or not, the corresponding amplitudes
should be summed.)  The contribution
from the topologically trivial sector (no black hole), which he had
previously neglected, is sufficient to restore unitarity (Hawking 2005).
Hawking's arguments so far are schematic
and will likely be followed up by more detailed computations.  In any
case, it is clear that his arguments do not yet explain where the standard
semiclassical argument breaks down, if it does.

Turn lastly to loop quantum gravity.  Here, the first key result
relevant to information loss is that the singularity inside
Schwarzschild black holes is resolved (Ashtekar \& Bojowald 2005b).
The analysis is similar to earlier analyses of the resolution of
cosmological singularities (Bojowald 2001, Ashtekar et.\ al.\ 2003).
In the Schwarzschild case the theory that is quantized is a
a finite-dimensional, minisuperspace model rather than the full,
infinite-dimensional theory.
However, the particular quantization chosen (which is unconventional)
is motivated by considerations of the loop quantum
gravity program on the full phase space.  With this quantization, the
Hamiltonian constraint reduces to a finite difference equation, which
can be thought of as describing a discrete time evolution.  For
appropriate solutions of the difference equation, operators
describing the geometry are well defined at the singularity,
where they diverge classically.  While one might object to the
truncation of the phase space it is plausible
that similar results might hold in the full theory.

Ashtekar \& Varadarajan (2005) also analyze the more complicated model
of the two-dimensional CGHS black hole (Callan, Giddings, Harvey \&
Strominger 1992), which incorporates Hawking radiation and backreaction.
The key results here are (i) the singularity is again resolved, and
(ii) there exists a region of the spacetime surrounding what would be
the singularity where quantum gravitational
effects are strong and a classical geometry does not exist.
Before this region, and again after it, a
semiclassical approximation is valid.
In particular there are no baby universes, and no superpositions of
macroscopically different geometries at late times.
Ashtekar \& Bojowald (2005a) argue that
these conclusions should also be valid in four-dimensions
If this is the case, then the fact that the singularity is
resolved should imply that black-hole formation and evaporation is
unitary, since there are no singularities in the quantum theory.

How is this loop quantum gravity scenario consistent with the standard
semiclassical argument for information loss?  The answer to this
question is not yet completely clear.  Ashtekar and Bojowald (2005a)
speculate that the amount of mass that passes through the singularity
(the mass contained inside the surface $S$ discussed above) could be
much larger than the
Planck mass.  This would require the semiclassical solution that describes
the evaporation outside the black hole to break down before the Planck
regime is reached, or to have properties other than those usually
assumed (for example, an apparent horizon area of order the Planck area at the
same retarded time as a Bondi mass much greater than the Planck mass;
Ashtekar 2005).  This option cannot be ruled out since a
complete and detailed computation of the semiclassical solution has
not yet been carried out in four dimensions; however it seems unlikely
given the two-dimensional computations that have been performed.
If it is true that the mass that passes through the singularity is
much larger than the Planck mass, then the semiclassical argument for
information loss can be evaded.

One of the appealing features of the loop quantum gravity
computations, in my view, is that they are sufficiently explicit, direct and
local that it should be possible with additional computations to pin
down where and why the semiclassical argument fails.
By contrast, the Euclidean quantum gravity approach restricts itself
to computing asymptotic observables; it is difficult using only these
observables to try to understand why the local semiclassical arguments
fail.  Similarly, in string theory the most detailed understanding of
the evaporation process is in terms of the dual description in the
boundary conformal field theory; it is not easy to translate this into
a detailed local understanding of the evaporation process (although
see Lowe \& Thorlacius 1999).

The most satisfying resolution of the information loss paradox, in my
view, would be an explanation of why the semiclassical theory breaks
down earlier than naively expected.  Ideally this explanation would
use only minimal assumptions about quantum gravity, and would be
applicable to all three of the approaches to quantum gravity discussed
here.  The recent loop quantum gravity computations might be a step in
this direction.

\bigskip

{\bf References:}

Ashtekar, A., 2005, private communication.
\hfill\break
Ashtekar, A. \& Bojowald, M., 2005a, Class. Quant. Grav. {\bf 22}, 3349
(also
\htmladdnormallink{gr-qc/0504029}{http://arXiv.org/abs/gr-qc/0504029}).
\hfil\break
Ashtekar, A. \& Bojowald, M., 2005b, {\it Quantum geometry and the
  Schwarzschild singularity},
  \htmladdnormallink{gr-qc/0509075}{http://arXiv.org/abs/gr-qc/0509075}.
\hfil\break
Ashtekar, A., Bojowald, M. \& Lewandowski, L., 2003,
Adv. Theor. Math. Phys. {\bf 7}, 233.
\hfil\break
Ashtekar, A. \& Varadarajan, M., 2005, in preparation.
\hfil\break
Bojowald, M., 2001, Phys. Rev. D {\bf 64}, 084018
\hfil\break
Bose, S., Parker, L. \& Peleg, Y., 1996, Phys. Rev. D {\bf 53}, 7089.
\hfil\break
Callan, C.\ G., Giddings, S.\ B., Harvey, J.\ A. \& Strominger, A.,
1992, Phys. Rev. D {\bf 45}, R1005.
\hfil\break
Gambini, R., Porto, R. \& Pullin, J., 2004, Phys. Rev. Lett. {\bf 93}, 240401.
\hfil\break
Hawking, S.W., 1976, Phys. Rev. D {\bf 14}, 2460.
\hfil\break
Hawking, S.W., 2004, presentation at 17th International Conference on
General Relativity and Gravitation, Dublin, Ireland; transcript
available at \\
\htmladdnormallink{http://pancake.uchicago.edu/\~{}carroll/hawkingdublin.txt}
{http://pancake.uchicago.edu/~carroll/hawkingdublin.txt}.
\hfil\break
Hawking, S.W., 2005, {\it Information Loss in Black Holes},
\htmladdnormallink{hep-th/0507171}{http://arXiv.org/abs/hep-th/0507171}.
\hfil\break
Horowitz, G. \& Maldacena, J., 2004, JHEP {\bf 008}, 0402
(also \htmladdnormallink{hep-th/0310281}{http://arXiv.org/abs/hep-th/0310281}).
\hfil\break
Jacobson, T., Marolf, D. \& Rovelli, C., 2005, {\it Black hole
entropy: inside or out?},
\htmladdnormallink{hep-th/0501103}{http://arXiv.org/abs/hep-th/0501103}.
\hfil\break
Lowe, D.A., Polchinski, J., Susskind, L., Thorlacius, L. \& Uglum, J.,
1995, Phys. Rev. D {\bf 52}, 6997 (1995).
\hfil\break
Lowe, D.A. \& Thorlacius, L., 1999, Phys. Rev. D {\bf 60}, 104012.
\hfil\break
Page, D.N., 1976, Phys. Rev. D {\bf 14}, 3260.
\hfil\break
Page, D.N., 1994, {\it Black Hole Information},
in ``Proceedings of
the 15th Canadian Conference on General Relativity and Relativistic
Astrophysics'', Eds. R.B. Mann and R.G. McLenaghan, World Scientific,
Singapore (also
\htmladdnormallink{hep-th/9305040}{http://arXiv.org/abs/gr-qc/9305040}).
\hfil\break
Preskill, J., 1993, {\it Do black hole destroy information?},
in
``International symposium on black holes, membranes, wormholes and
superstrings'', World Scientific, River Edge, NJ
(also
\htmladdnormallink{hep-th/9209058}{http://arXiv.org/abs/hep-th/9209058}).
\hfil\break
Strominger, A. 1994, {\it Les Houches lectures on black holes},
\htmladdnormallink{hep-th/9501071}{http://arXiv.org/abs/hep-th/9501071}.
\hfil\break
Thorlacius, L. 2004, {\it Black Holes and the Holographic Principle},
\htmladdnormallink{hep-th/0404098}{http://arXiv.org/abs/hep-th/0404098}.

\vfill
\eject

\section*{\centerline {
Gravity Probe B mission ends}}
\addcontentsline{toc}{subsubsection}{\it  
Gravity Probe B mission ends, by Bob Kahn}
\begin{center}
 Bob Kahn,  Stanford University
\htmladdnormallink{cmw-at-howdy.wustl.edu}
  {mailto:cmw@howdy.wustl.edu}
\end{center}

On August 15, 2005, Gravity Probe B completed 352 days of
science data collection 
and conducted a series of
final instrument calibration tests before the liquid helium in the Dewar was
exhausted around Labor Day. At that point, the main focus of GP-B
shifted from mission operations to data analysis. The 
scientific analysis work will require over a year to complete, followed by
up to six months of preparing and submitting scientific papers to major
scientific journals. This process will culminate in the announcement and
publication of the results, now anticipated to occur around April 2007. 
Thus it seems appropriate to provide an overview of what
is involved in the GP-B data analysis process.

Recall that GP-B consists fundamentally of four spinning gyroscopes and a
telescope.
Conceptually the experimental procedure is simple: At the beginning of
the experiment, we point the science telescope on-board the spacecraft at
a guide star, IM Pegasi, and we electrically nudge the spin axes of the
four gyroscopes into the same alignment parallel to the telescope axis. 
Then, over the course of a year, as
the spacecraft orbits the Earth some 5,000 times while the Earth makes one
complete orbit around the Sun, the four gyros spin undisturbed---their spin
axes influenced only by the relativistic warping and twisting of spacetime.
We keep the telescope pointed at the guide star using attitude control
thrusters on the spacecraft, and each orbit, we record
the cumulative size and direction of the angle between the gyroscopes' spin
axes and the telescope. According to the predictions of Einstein's general
theory of relativity, over the course of a year, an angle of 6.6 arcseconds
should open up in the plane of the spacecraft's polar
orbit, due to the warping of
spacetime by the Earth (geodetic effect), 
and a smaller angle of 0.041 arcseconds should open
up in the direction of Earth's rotation due to the Earth dragging its local
spacetime around as it rotates (Lense-Thirring effect).

In reality, what goes on behind the scenes in order to obtain these gyro
drift angles is a complex process of data reduction and analysis that
will take the GP-B science team more than a year to bring to completion. 
We continuously
collect data during all scheduled telemetry passes with ground stations and
communications satellites, and this telemetered data is stored--in its raw,
unaltered form--in a database at the GP-B Mission Operations Center at
Stanford University.
This raw data is called Level 0 data. The GP-B spacecraft is capable of
tracking some 10,000 individual values, but we only capture about 1/5 of
that data. The Level 0 data includes a myriad of status information on all
spacecraft systems in addition to the science data, all packed together for
efficient telemetry transmission. The first data reduction task is to
extract all of the individual data components from the Level 0 data and
store them in the database with mnemonic identifier tags. These tagged data
elements are called Level 1 data. We then run a number of algorithmic
processes on the Level 1 data to extract around 500 data elements that will be
used for science data analysis; this is Level 2 data. While Level
2 data include information collected during each entire orbit, the science
team generally only uses information collected during the 
portion of each orbit when the telescope is locked onto the guide
star. We do not use for science 
any gyroscope or telescope data collected during that
portion of each orbit when the spacecraft is
behind the Earth, eclipsed from a direct view of the guide star.

If there were no noise or error in our gyro readouts, and if we had
known the exact calibrations of these readouts at the beginning of the
experiment, then we would only need two data points--a starting point for
the gyroscope orientations and an
ending point. However, since we are determining the exact readout
calibrations as part of the experiment, collecting all of the data points in
between enables us to determine these unknown variables. 

Another important point is that the electronic systems on-board the
spacecraft do not read out angles. Rather, they read out voltages, and by
the time these voltages are telemetered to Earth, they have
undergone many conversions and amplifications. Thus, in addition to the
desired signals, the GP-B science data includes a certain amount of random
noise, as well as various sources of interference. The random noise averages
out over time and is not an issue. Some of what appears to be regular,
periodic interference in the data is actually important calibrating signals
that enable us to determine the size of the scale factors that accompany the
science data.  For example, the
orbital and annual aberration of the starlight from IM Pegasi is used as a
means of calibrating the gyro readout signals.  As the telescope is
continually reoriented to track the apparent position of the guide star, an
artificial, but accurately calculable, periodically varying
angle between the gyros and
the readout devices is introduced.  This allows the precise measurement
of the voltage-to-angle scale factor.
Measurement of this factor is optimized by a full year's worth of annual
aberration data.

Finally, there is one more very important factor that must be addressed in
calculating the final results of the GP-B experiment. We selected IM Pegasi,
a star in our galaxy, as the guide star because it is both a radio source
and it is visually bright enough to be tracked by the science telescope
on-board the spacecraft. Like all stars in our galaxy, IM
Pegasi moves relative to the solar system
because of its local gravitational environment and because of
galactic rotation.
Thus, the GP-B science telescope is tracking a moving
star, but the gyros are unaffected by the star's so called proper motion; their
pointing reference is IM Pegasi's position at the beginning of the
experiment. Thus, each orbit, we must subtract out the telescope's angle of
displacement from its original guide star orientation so that the angular
displacements of the gyros can be related to the telescope's initial
position, rather than its current position. The motion of IM Pegasi
with respect to a distant quasar has been measured with extreme precision
over a number of years using Very Long Baseline
Interferometry (VLBI) by a team at the Harvard-Smithsonian Center for
Astrophysics (CfA) led by Irwin Shapiro, in collaboration with
astrophysicist Norbert Bartel and others from York University in Canada and
French astronomer Jean-Francois Lestrade. However, to ensure the integrity
of the GP-B experiment, we added a "blind" component to the data analysis by
insisting that the CfA withhold the proper motion data that will enable us
to pinpoint the orbit-by-orbit position of IM Pegasi until the rest of our
data analysis is complete. Therefore, the actual drift angles of the GP-B
gyros, the quantities that are to be compared with the predictions of
general relativity,
will not be known until the very end of the data analysis process.

For additional information about the GPB project, go to the website
\htmladdnormallink
{http://einstein.stanford.edu/}
{http://einstein.stanford.edu/}.
\vfill
\eject

\section*{\centerline 
{ 6th Edoardo Amaldi Meeting}}
\addtocontents{toc}{\protect\medskip}
\addtocontents{toc}{\bf Conference reports:}
\addcontentsline{toc}{subsubsection}{\it  
6th Edoardo Amaldi Meeting, by
Matthew Benacquista}
\parskip=3pt
\begin{center}
Matthew Benacquista,
Montana State University-Billings
\htmladdnormallink{benacquista-at-msubillings.edu}
{mailto:benacquista@msubillings.edu} \end{center}

The 6th Edoardo Amaldi Conference on Gravitational Waves was held at
Bankoku Shinryoukan in Okinawa, Japan on June 19th to 24th. It was
attended by approximately 200 participants. The maturity of the field of
gravitational wave detection was reflected in the two sessions devoted
to current detectors, three sessions on data analysis, and two sessions
on research and development for advanced detectors. In addition to an
overview session, there were also one session devoted to gravitational
wave sources and one to LISA.

The reports on the status of current detectors stressed the approach to
design sensitivity for the interferometers, remarkable improvements in
duty cycle and bandwidth for the bar detectors, and the impending
science run for the spherical resonant detector miniGrail.

TAMA reported the accumulation of over 3000 hours of data and the
beginning of automated, crewless runs in late 2003. The increase in
sensitivity has been up to four orders of magnitude at some frequencies
in the sensitivity band and the detector is approaching design
sensitivity. This has given TAMA sufficient sensitivity to observe the
entire galaxy. The planned installation of the seismic attenuation
system will offer improved sensitivity for the next planned science run
in early 2006. LIGO reported that all three interferometers are within
roughly a factor of 2 of design sensitivity and that the installation of
a preisolator system for the Livingston interferometer has significantly
reduced the impact of the local anthropogenic seismic disturbances.
Virgo has been commissioning since November, 2003 and now, after a
reduction in noise of $10^4$ in one year, it is less than 2 orders of
magnitude above design sensitivity at frequencies around 200 Hz and
higher. Although the low fequency sensitivity is still several orders of
magnitude above design sensitivity, continued noise hunting is expected
to bring this level down as well. GEO continues in its role as both
detector and prototype as its sensitivity has improved to within an
order of magnitude of design in narrowband tuning at 1 kHz.

ALLEGRO has achieved a 95\% duty cycle since May, 2004. It has run with
3 rotations during S4 of LIGO to search for a stochastic signal.
Otherwise, it was run in parallel with the European bars as part of
IGEC-2. Hardware improvements are planned for ALLEGRO in the near
future. AURIGA reports a resolution of a problem of spurious peaks that
had plagued it in early 2004. It has now achieved design sensitivity and
a 95\% duty cycle since May, 2005. They report remarkably stable,
Gaussian noise over its broad 100 Hz bandwidth. The Nautilus/Explorer
bars have now been running in coincidence since March, 2004 with an 85\%
duty cycle and a strain sensitivity of $\sim 3.5 \times 10^{-19}$ near
920 Hz. Finally, miniGRAIL anticipates the first science run in early
2006 after improvements in the sphere and the transducer.

The data analysis sessions yielded a variety of talks covering ongoing
analysis on current data, as well as proposed analysis schemes for
current and future data. Of particular interest from the talks on
ongoing analysis was the results from Explorer/Nautilus `03 science run
which has ruled out the suggestion of an excess of coincidences in the
`01 science run. Although still too large to be of astrophysical
interest, the upper limits on continuous wave strain amplitudes of $h
\sim 6 \times 10^{-23}$ from LIGO are approaching an order of magnitude
above astrophysically reasonable values. The reports on binary
coalescence rates from LIGO are still far too large, but the scientific
reach of more than 20 Mpc from S4 is beginning to become interesting.
Improvements in the search for stochastic backgrounds with LIGO/ALLEGRO
have also begun to approach limits achieved by big bang nucleosynthesis.
A number of talks focused on progress in the development of joint data
analysis projects between detectors. This is of significant interest as
it heralds the development of a worldwide network of detectors. In
particular, the outline of a new agreement for IGEC-2 is expected to
produce significant improvements in performance of a search for events.
Finally, an implementation of maximum entropy techniques was presented
that may have implications for improved angular resolution of both
ground-based and space-based interferometers.

The sessions on advanced detectors included summaries of the plans and
progress for improving existing detectors as well as designs for
completely new detectors. For most of the existing interferometers, the
anticipated improvements are expected to come from better seismic
control at low frequencies, higher power lasers at high frequencies, and
lower thermal noise in the mid-band. Advanced LIGO is now planning for
installation at LLO in 2010 with commissioning runs expected to begin in
2012. Improved Virgo is still in the creative stage but is studying
similar solutions to improve the sensitivity. Progress is reported on
LCGT which will achieve its improved sensitivity by going underground
and going cryogenic. They have begun testing mirrors for thermal noise
in cryogenic conditions, and have measured significantly improved
seismic noise at the planned site. The first underground cryogenic
interferometer, CLIO, reports that construction is ongoing with the
completion of the vacuum system and significant parts of the cryogenic
systems.

Overall plans for improving the existing bar detectors have focused on
increasing sensitivity by going to lower temperature, and increasing
bandwidth by improving transducers and amplifiers. Most reports focused
on new transducer technology and double SQUID amplifiers. Progress is
being made on the Brazilian spherical detector. Work is ongoing with the
DUAL detector, which anticipates achieving a wide bandwidth by nesting
two bars with different resonant frequencies. Progress is reported on
the development of transducers for this arrangement.

Research and development in advanced detectors has covered a number of
areas for improving sensitivity. Work on squeezed light has continued
with a goal of beating the quantum limit. Improvements in high power
lasers were reported, along with the necessary work on improved thermal
noise properties and coatings for mirrors. Alternate beam shapes (flat
top profiles) have also been investigated.

The session devoted to LISA was more devoted to LISA Pathfinder, the
upcoming technology demonstrator with a planned launch in 2009. The
planning and design of LTP is almost completed and testing indicates
that it approaches or exceeds the requirements. Construction is now
underway. The ongoing design and testing of the LISA gravitational
reference systems at Trento using their torsion pendulum is also
approaching the LTP requirements and the capabilities of the pendulum
are now limiting further testing. In addition to LISA, there were talks
about possible designs for the gravitational reference system of the
proposed LISA follow-on mission (BBO). An update on DECIGO, the proposed
Japanese deciHertz detector, reports progress and a possible launch date
of 2022. Details of the model dependence of the probable Galactic white
dwarf binary background were presented as well as a discussion of
testing the non-linear aspects of gravity with LISA.

The session on sources included an overview of expected sources for
ground-based detectors, focusing mainly on compact object inspirals. Of
particular note was a seven-fold increase in the expected event rate for
neutron star inspirals based upon the observation of J0737-3039.
Additional work on numerical simulation of waveforms using Whisky was
reported. This will be of value to both ground-based and space-based
interferometers. A detailed study of a simulation of the white dwarf
binary population that included mass transfering systems was presented
as a background signal for LISA. Finally, at the high frequency end of
the spectrum, the potential for Low-Mass X-ray Binaries to excite
f-modes may provide gravitational wave signals at high frequencies.

This was an exciting meeting as several detectors are now operating and
producing data, while at the same time work is progressing toward the
implementation of second generation detectors. Meanwhile, a number of
collaborative efforts are underway that promise to usher in the era of a
world-wide network for the detection of gravitational radiation. The
next Amaldi meeting, to be held in Sydney, should be equally exciting as
much of the progress reported here will be producing results in two
years.

\vfill\eject

\section*{\centerline {
Workshop on Numerical Relativity,}\\\centerline{ Banff International Research
Station}}
\addcontentsline{toc}{subsubsection}{\it  
Workshop on Numerical Relativity, BIRS, by Carsten Gundlach}
\begin{center}
Carsten Gundlach, University of Southampton
\htmladdnormallink{C.Gundlach-at-maths.soton.ac.uk}
  {mailto:C.Gundlach@maths.soton.ac.uk}
\end{center}

The workshop was organized Doug Arnold, Matt Choptuik,
Luis Lehner, Randy LeVeque and Eitan Tadmor, with the purpose of
bringing together researchers in GR working numerically and
analytically. 20 invited half-hour talks were given over 4 days, with
plenty of time for discussions between talks, over meals, and in the
evening. 

The BIRS page on the programme can be found on

\htmladdnormallink
{http://www.pims.math.ca/birs/}
{http://www.pims.math.ca/birs/}

and Matt Choptuik's page including PDF files of talks is 

\htmladdnormallink
{http://bh0.physics.ubc.ca/BIRS05/}
{http://bh0.physics.ubc.ca/BIRS05/}

To complement this, I shall highlight only a few of the talks.

In the 1990s, some researchers were concentrating on obtaining physics
insight from effectively 1+1 dimensional problems: what cosmological
spacetimes with two commuting Killing vectors can tell us about the
nature of generic singularities (Berger and collaborators), and what
we can learn about cosmic censorship from spherical collapse (Choptuik
and students). More ambitious, axisymmetric or 3D, work confronted
overlapping problems hard to disentangle in the low resolution
available in 3D. In particular, instabilities already present in the
continuum problem were not clearly distinguished from those added at
the discretization stage. The Banff meeting showed that now at least
we have a clearer view of the problems facing us.

3+1 approaches need to start from a well-posed initial-boundary value
problem in the continuum, with boundary conditions that are compatible
with the constraints. Well-posedness can be proved by energy methods,
based on a symmetric hyperbolic form of the field equations. {\em
Olivier Sarbach} drops the energy estimate based on the symmetrizer in
favor of a ``physical'' energy plus a constraint energy. The
remaining ``gauge'' energy is estimated separately using elliptic
gauge conditions. This intuitively appealing programme has been
completed for electromagnetism, although the gauge seems a bit
restrictive. Work with Nagy is under way on general relativity. By
contrast {\em Oscar Reula} emphasized that strong hyperbolicity is
often enough. He could prove that whenever a first-order system
subject to constraints is strongly hyperbolic (eg the BSSN
formulation) then so is the associated constraint evolution
system. {\em Heinz Kreiss} surprised some of his disciples in the
numerical relativity community by also stressing that energy methods
are too limited. In a series of examples, he proposed a general
approach based on reducing initial-boundary value problems to
half-space problems with frozen coefficients and analyzing the
dependence of each Fourier mode on its initial and boundary data.

On the numerical methods front, {\em Manuel Tiglio} reported on
collaborative work to discretize systems of first-order strongly
hyperbolic equations on multiple touching patches (for example 6 cubes
to form a hollow sphere), using summation by parts and penalty
methods. Their animations of toy problems looked very impressive, and
the whole technology will be available as a general tool through the
Cactus infrastructure. {\em Michael Holst} and {\em Rick Falk} gave
review talks on finite elements for both elliptic and evolution
equations. This is promising for nontrivial domains, but has not yet
been applied to numerical relativity.

Other talks showed what 3D simulations can do. {\em David Garfinkle}
reported on simulations of cosmological singularities without any
symmetries on $T^3$. The key elements of his approach are the use of
inverse mean curvature flow slicing ($\alpha=1/K$) and a tetrad and
connection formulation used successfully by Uggla and coworkers in
analytical studies. His results are compatible with the BKL
conjecture, although soon the resolution becomes too low to follow the
development of ever more decoupled Bianchi IX regions. {\em Thomas
Baumgarte} summarized the state of the art in binary neutron star
simulations by himself and others, notably Masaru Shibata. There seems
to be no real showstopper for such simulations. Rather what is needed
now is more resolution, and the modelling of physical phenomena such
as neutrinos, viscosity, and magnetic fields. Interesting results
include the formation in binary mergers of a hot neutron star held up
only by differential rotation, and expected to collapse later.

The most noted talk of the meeting was that of {\em Frans Pretorius}
giving preliminary results on binary black hole mergers using harmonic
coordinates. His simulations no longer seem to be limited by
instabilities, but rather by computer power and time, and by
unphysical initial data (there is evidence that his initial data are
very far from circular inspiral data). The key ingredients seem to be
the following: a working 3D AMR code on still massive computers,
compactification of the Cartesian spatial coordinates (that is, at
$i_0$) together with damping of outgoing waves, modified harmonic
coordinates, and a damping of the harmonic gauge constraint through
lower order friction terms (Gundlach). Generalized
harmonic gauge (Friedrich) is $\Box x^\mu=H^\mu$, where the gauge
source functions $H^\mu$ are treated as given functions. Pretorius
makes $H^0$ obey a wave equation $\Box H^0\sim \alpha-1$, which
prevents the lapse from collapsing without affecting the
well-posedness. This works less well for critical collapse.

\vfill\eject

\section*{\centerline {
8th Capra Meeting on Radiation Reaction}}
\addcontentsline{toc}{subsubsection}{\it  
8th Capra Meeting on Radiation Reaction, by Leor Barack}
\begin{center}
Leor Barack, University of Southampton
\htmladdnormallink{L.Barack-at-maths.soton.ac.uk}
  {mailto:L.Barack@maths.soton.ac.uk}
\end{center}

The {\it Capra} meetings are annual gatherings of theorist
working on the problem of radiation reaction in General Relativity.
Previous meetings were held in Southern California (in 1998,
at a ranch donated to Caltech by former Caltech alumnus, Hollywood
director Frank Capra), Dublin, Caltech, Potsdam (Germany), Penn State,
Kyoto, and Brownsville (Texas). The Capra meetings focus on the study
of radiation reaction and self interaction (``self force'') in curved
spacetime, especially in the context of the general relativistic two-body
problem. Recent interest in this problem has been strongly driven by
the prospects of detecting gravitational waves from inspirals of compact
objects into supermassive black holes, using the planned LISA detector.
There has been a remarkable progress over the last few years on the problem
of modeling such inspirals, much through the work of the Capra community.

This year's meeting---the eighth in the series---took place July 11-14th
in Abingdon (Oxfordshire), UK. Organized jointly by the Southampton Relativity
Group (N. Andersson, L. Barack, K. Glampedakis) and the Centre for Fundamental
Physics at RAL (R. Bingham), the meeting was hosted at RAL's
{\it Cosener's house}---a conference facility situated at a picturesque
Thames side in the grounds of the medieval Abbey of Abingdon.
41 researchers attended the meeting this year, representing universities
in 8 countries. The scientific agenda included 1-hour contributed talks,
with a significant portion of the time reserved for questions and discussion.
The topics covered ranged from issues in the fundamental
formulation of self forces in curved spacetime, through advances in black hole
perturbation theory, to actual computations of the self force effect in black
hole orbits. The last part of the meeting focused on detection aspects
and data-analysis strategies.

The meeting opened with a user-friendly ``self-force primer''
by Steve Detweiler (Gainesville). Eric Poisson (Guelph) followed with a talk
on the metric of tidally distorted black holes, and Warren Anderson (Milwaukee)
presented results from an analytic computation of the local tail contribution
to the gravitational self force.
Leor Barack (Southampton) concluded the first day with a talk on the
Lorenz-gauge formulation of black hole perturbation theory, with
applications to self force calculations. In the second day of the
meeting, Dong-Hoon Kim (Gainesville) presented his calculation of the
``regularization parameters'' necessary for calculating the gravitational
self-force on particles in circular orbits around Schwarzschild black holes.
Steve Detweiler then showed how these can be used to deduce some
gauge-invariant self-force effects for such orbits.
Hiroyuki Nakano (Osaka City University) next explained how to derive the
regularization parameters for generic orbits in Schwarzschild, and was
followed by Waratu Hikida (Kyoto), who presented first results from
calculations of the scalar self-force for eccentric orbits.
Sophiane Aoudia (Nice) discussed the calculation of the regularization
parameters in the case of a particle plunging radially into a
black hole. A fully-numerical approach to the problem was
introduced by Carlos Sopuerta (Penn State), who presented first results
from time-domain finite-element numerical simulations of extreme-mass-ratio
inspirals.

The third day of the meeting was opened by Carlos Lousto (Brownsville)
with a talk on recent progress on the problem of reconstructing the
metric perturbation in black hole spacetimes. This was followed up by
Bernard Whiting (Gainesville), who discussed further ideas on how one might
go about reconstructing the metric perturbation in Kerr spacetime, and by
Larry Price (Gainesville) who presented a useful Maple toolkit for carrying
out calculations in the GHP approach to Kerr perturbations. Eran Rosenthal
(Guelph) wrapped up this part of the meeting with a talk on regularization
of the second-order self force.

In opening the final, ``detection aspects'' part of the meeting, Jonathan
Gair (Cambridge) presented work done to develop a set of ``quick and dirty''
approximate waveforms for extreme-mass-ratio inspirals, which had
been used to scope out data-analysis issues. Norichika sago (Osaka)
reviewed the underlying formalism for an alternative set of approximate
waveforms, based on a form of adiabaticity assumption and a
time-averaging procedure. A numerical code for calculating such adiabatic
waveforms was presented by Steve Drasco (Cornell) at the beginning of the
last day of the meeting. He argued that such waveforms are likely to be
sufficiently accurate to enable detection of extreme-mass-ratio inspirals
with LISA, but perhaps not accurate enough to allow full extraction of
system parameters. Next, Kostas Glampedakis (Southampton) presented
a formalism that could be used to quantify deviations from Kerr geometry,
as encoded in the waveforms from extreme-mass-ratio inspirals.
Jonathan Gair then discussed the challenges and work done on developing
data-analysis techniques for searching over inspirals in the LISA data
stream. Gareth Jones (Cardiff) described a search method based on
an algorithm for identifying ``clusters and ridges'' on a time-frequency
spectrogram. Alberto Vecchio and Alexander Stroeer (Birmingham)
presented a data analysis scheme for detecting interacting white-dwarf
binaries, with lessons for detection of extreme-mass-ratio inspirals.
Finally, Charles Wang (Aberdeen) discussed the consequences of photons
interacting with gravitational waves from compact objects.

An electronic version of all talks given in the meeting is available
online at

\htmladdnormallink
{http://www.sstd.rl.ac.uk/capra/}
{http://www.sstd.rl.ac.uk/capra/}

The organizers acknowledge the generous financial support of RAL and
the IOP Gravitational Physics Group.

\vfill\eject

\section*{\centerline {
Theory and experiment in quantum gravity}}
\addcontentsline{toc}{subsubsection}{\it  
Theory and experiment in quantum gravity, by Elizabeth Winstanley}
\begin{center}
Elizabeth Winstanley, University of Sheffield
\htmladdnormallink{E.Winstanley-at-sheffield.ac.uk}
  {mailto:E.Winstanley@sheffield.ac.uk}
\end{center}

Quantum gravity is a wide-ranging subject, with many different
theoretical approaches and the exciting possibility of probing quantum
gravity phenomenology in the near future.  The aim of this meeting was
to give an overview of current research in at least some areas of
quantum gravity, both theoretical and experimental.  The talks were
pedagogical in nature and accessible to PhD students, and the meeting
informal, with plenty of time for discussion.  The meeting was held in
the Ogden Centre for Fundamental Physics, University of Durham, on
July 7-8, and organized by Ruth Gregory (Durham) and Elizabeth
Winstanley (Sheffield).

The first day began with three talks on theoretical approaches to
quantum gravity.  Fay Dowker (Imperial College London) gave an
introduction to the concept of causal sets and the new construction of
swerves for particle paths on a causal set; John Barrett (Nottingham)
reviewed current research in spin foams and the latest developments in
$3+0$-dimensional quantum gravity; and Bernard Kay (York) introduced
the theory of quantum field theory in curved space, and its
applications to the Casimir effect and black hole radiation.  The
second session of the first day was devoted to higher dimensions and
branes.  Tony Padilla (Oxford) explained the particular features of
brane world gravity, and focussed on the idea of braneworld
holography; and Christos Charmousis (Orsay) covered higher derivative
(particularly Lovelock) gravity, and its importance for
$4+N$-dimensional spacetimes.  The first day ended with a talk by
Panagiota Kanti (Durham) on the Hawking radiation of
higher-dimensional brane black holes.

The second day began with cosmology: Ian Moss (Newcastle-upon-Tyne)
spoke about quantum effects in brane cosmology, and the role of
boundary conditions in Horava-Witten theory.  Ivonne Zavala (Boulder)
brought us up-to-date with developments in brane inflation in string
theory.  Then the emphasis changed to experimental areas: Joy
Christian (Oxford) explored the forthcoming experimental possibilities
of probing the Planck scale with cosmogenic neutrinos, and Giles
Hammond (Birmingham) introduced the new experiments testing the
Casimir force and the inverse square law at short range.

This two-day meeting was attended by over 60 people, including many
graduate students.  The organizers would like to thank the
Mathematical \& Theoretical Physics and Gravitational Physics Groups
of the Institute of Physics for financial support.

\end{document}